\documentclass[aps,showpacs,amsmath,amssymb,twocolumn,floatfix,pra]{revtex4}
\usepackage[dvips]{graphicx}
\input{epsf}
\begin{document}
\newcommand{\newc}{\newcommand}
\newc{\beq}    {\begin{equation}}
\newc{\eeq}    {\end{equation}}
\newc{\beqa}    {\begin{eqnarray}}
\newc{\eeqa}    {\end{eqnarray}}
\newc{\no}    {\\ \nonumber}
\def\PL{{\em Phys. Lett.}}
\def\PLA{{ Phys. Lett.} { A} }
\def\NPB{{\em Nucl. Phys.} { B} }
\def\PRL{{ Phys. Rev. Lett. }}
\def\PRB{{ Phys. Rev.} { B} }
\def\PRA{{ Phys. Rev.} { A} }

\title{ A model of deterministic detector with dynamical decoherence}

\author{}
\author{Jae Weon Lee$^{(a)}$,  Dmitri V. Averin$^{(b)}$,
Giuliano Benenti$^{(c)}$ and 
Dima L. Shepelyansky$^{(a)}$}
\affiliation{$^{(a)}$Laboratoire de Physique Th\'eorique, 
UMR 5152 du CNRS, Univ. P. Sabatier, 31062 Toulouse Cedex 4, France\\
$^{(b)}$Department of Physics, University of Stony Brook, SUNY, Stony Brook, NY 11794 \\
$^{(c)}$Center for Nonlinear and Complex Systems, Universit\`a degli 
Studi dell'Insubria and Istituto Nazionale per la Fisica della Materia, 
Unit\`a di Como, Via Valleggio 11, 22100 Como, Italy
}

\date{January 26, 2005}

\begin{abstract}
We discuss a deterministic model of detector coupled 
to a two-level system (a qubit). The detector is a quasi-classical 
object whose dynamics is described by the kicked rotator
Hamiltonian. We show that in the regime of quantum chaos
the detector acts as a chaotic 
bath and induces decoherence of the qubit.
We discuss the dephasing and relaxation rates and demonstrate
that several features of Ohmic baths can be reproduced by 
our fully deterministic model.
Moreover, we show that, for strong enough qubit-detector 
coupling, the dephasing rate is given by the rate of 
exponential instability of the detector's dynamics, that is,
by the Lyapunov exponent of classical motion. 
Finally, we discuss the measurement in the regimes of 
strong and weak qubit-detector coupling. For the case of strong coupling
the detector performs a measurement of the up/down state of the qubit. 
In the case of weak coupling, due to chaos, the dynamical 
evolution of the detector is strongly sensitive to the state of 
the qubit. However, in this case it is unclear how to extract a 
signal from any measurement with a coarse-graining in the phase
space on a size much larger than the Planck cell.
\end{abstract}

\maketitle

\section{Introduction}

The long-standing problem of quantum measurement 
\cite{braginsky,giulini,pascazio} has recently gained 
a renewed interest due to its relevance for quantum 
information processing. 
Indeed, one of the requirements for the physical 
implementation of quantum computation is the ability to 
readout a single two-level quantum system (qubit).
This problem has been solved in ion-trap quantum computation
\cite{blatt,wineland} using quantum jump detection.
In solid-state implementations, the single-qubit measurement 
is very challenging and has been widely discussed 
\cite{gurvitz,korotkov,averin,milburn,berman,ortiz,fazio,schon}. 
Moreover, various readout schemes have been
experimentally realized \cite{saclay,mooij,nakamura,contmeas}. 

A detector can be seen as a complex quasi-classical object 
coupled to a quantum system. It is therefore interesting 
to investigate the dynamical evolution of concrete 
system-detector models.
In this paper, we introduce and study a model in which 
the detector exhibits a complex quasi-classical dynamics 
and is coupled to a two-level quantum system. The detector
is described by the kicked rotator Hamiltonian \cite{felix}, so 
that its dynamics can be, in the classical limit, chaotic, integrable or 
with mixed phase space. An interesting feature of 
our model is that it can be, in principle, realized by means 
of cold atoms in optical lattices. Cold atoms exposed to 
periodic standing waves of light proved to be an
ideal testing ground to explore the quantum dynamics of nonlinear 
systems \cite{raizen,phillips} and, in particular, made possible 
to realize experimentally the quantum kicked rotator
\cite{raizen2,amman,delande,darcy}. Moreover, the sensitivity of 
the quantum dynamics to a control parameter has been 
recently measured for a classically chaotic system and 
suggested as a promising technique for precision measurements
\cite{darcy2}. 

It is clear that the process of quantum measurement 
involves a tradeoff between information gain and disturbance
\cite{nondemolition}. This means that the price to pay for 
the information to flow from the system to the detector
is the backaction by the detector. The backaction disrupts 
the coherent evolution of the quantum system. 
This entails a fundamental connection between decoherence
and quantum measurements.
In this context, the detector model introduced in this paper
can be seen as a model of chaotic environment 
\cite{chaoticbath}. We will 
show that indeed the coupling to the detector induces 
decoherence in the quantum system. We will also show that
the dynamical decoherence resulting from our fully deterministic
model has properties very similar to the decoherence induced
by a suitable, dissipative coupling to a non-deterministic 
environment. An interesting trait of our model is that the
decoherence rate is given, for strong enough coupling, 
by the Lyapunov exponent, namely by the rate of 
exponential instability of the detector's dynamics in the 
classical limit.

An interesting problem is what is the system-detector 
coupling strength required to measure a single qubit system.
We will show that, in our model, a strong coupling allows us 
to measure the up/down state of the qubit. On the other hand,
due to chaotic dynamics, also a weak coupling induces 
significantly different dynamical evolutions of the detector
in the case of qubit up or down state. 
However, it is not clear how to extract a detectable signal
from this difference.

The paper is organized as follows. In Sec.~II, we introduce 
our deterministic detector model and describe its possible implementation
by means of cold atoms in laser fields. In Sec.~III, we discuss the
decoherence of the dynamical system, induced by the the coupling
to the detector.
The results are compared with those of a phase damping map for the 
system's density matrix, describing the decoherence process in the 
quantum operations formalism. 
In Sec.~IV we analyze the dependence of the detector's response on the 
system-detector coupling strength. Our conclusions are given in 
Sec.~V. We also present an alternative derivation of the phase 
damping map, based on the master equation approach (Appendix A) and 
discuss the continuous limit of this map (Appendix B).

\section{Deterministic detector model}

We consider the interaction of a single qubit with a quantum 
kicked rotator.
The overall Hamiltonian $\hat{H}$ reads as follows:
\begin{equation}
\begin{array}{l}
{\displaystyle
\hat{H}=\hat{H}_s+\hat{H}_d+\hat{H}_{int}},\\
{\displaystyle
\hat{H}_s=\delta \hat{\sigma}_x},\\
{\displaystyle
\hat{H}_d=\frac{\hat{p}^2}{2}+
K \cos(\hat{\theta}) \sum_m\delta(t-m)},\\
{\displaystyle
\hat{H}_{int}= \epsilon_c \hat{\sigma}_z 
\cos(\hat{\theta})\sum_m \delta(t-m)}.
\end{array}
\label{Hamiltonian}
\end{equation}
Here $\hat{H}_s$ denotes the single-qubit Hamiltonian, 
$\hat{H}_d$ the detector's Hamiltonian 
(kicked rotator model \cite{felix})
and $\hat{H}_{int}$ the qubit-detector coupling. 
The Hamiltonian $H_s$ induces Rabi oscillations with
frequency $\omega_R=2\delta$ between the qubit levels 
$|0\rangle$ and $|1\rangle$ ($|0\rangle,|1\rangle$ denote
the eigenstates of $\hat{\sigma}_z$ corresponding to the
eigenvalues $+1$ and $-1$, respectively).
The detector is a particle moving in a periodic potential 
switched on/off instantaneously (kicks) at time intervals 
$\Delta\tau=1$. 
The time $t$ in Eq.~(\ref{Hamiltonian}) is measured 
in number of kicks.
We have $[\hat{p},\hat{\theta}]=-i\hbar$, where $\hbar$ is the 
effective dimensionless Planck constant.
The properties of the quantum kicked rotator are described in
\cite{felix}.
The interaction Hamiltonian is also kicked with
the same time period $\Delta\tau=1$. 
We note that the interaction Hamiltonian ($\propto \hat{\sigma}_z$)
does not commute with the system's Hamiltonian 
($\propto \hat{\sigma}_x$), namely we do not discuss nondemolition
measurements \cite{nondemolition}.
The unitary operator describing the evolution of the overall system 
(qubit plus detector) in one kick is given by  
\beq
\hat{U}=\exp\left[-i\frac{K+\epsilon_c 
\hat{\sigma}_z}{\hbar} \cos\hat{\theta}\right] 
\exp\left(-i\frac{ \hat{p}^2}{2\hbar}\right)
\exp(-i\delta\hat{\sigma}_x).
\eeq
Therefore, the effective kicking strength 
$K_{\rm eff}=K+\epsilon_c \sigma_z$
depends on the up or down state of the qubit. For this reason, the
kicked rotator can be, in principle, used as a detector.
We define $\epsilon=\epsilon_c/\hbar$ as the coupling strength
in units of $\hbar$.

The classical limit for the detector corresponds 
to $\hbar\to 0$, while keeping $K$ constant. 
The classical dynamics is integrable at $K=0$ and exhibits a
transition to chaos of the KAM (Kolmogorov-Arnold-Moser) type
when increasing $K$. The last invariant KAM torus is broken 
for $K\approx 1$ and the motion is fully chaotic (with no 
visible stability islands) when $K\gg 1$
\cite{chirikov79}.
In this paper, we will consider the detector as a quasi-classical 
object, that is, we will take $\hbar\ll 1$. 

In the numerical simulations discussed in this paper, we consider 
initial separable states $|\Psi\rangle=|\psi_s\rangle\otimes|\phi_d\rangle$,
where $|\psi_s\rangle=\alpha|0\rangle+\beta|1\rangle$ is a generic 
single-qubit state and $|\phi_d\rangle$ is a Gaussian
wave packet describing the initial state of the detector
(centered at $p=0$, $\theta=\pi$, with the uncertainties 
along $\theta$ and $p$ proportional to $\sqrt{\hbar}$).
We consider the kicked rotator dynamics 
on the torus $0\le \theta <2 \pi$, $-\pi \le p < \pi$.
The number $N_d$ of quantum levels describing the detector's dynamics 
is $N_d=2\pi/\hbar$. We consider $2^7\le N_d \le 2^{15}$, corresponding
to $4.91\times 10^{-2} \ge \hbar \ge 1.92\times 10^{-4}$.

To close this section, we would like to comment on the possibility
to implement, at least in principle, our model using cold atoms 
in a pulsed optical lattice created by laser fields.
For this purpose, the atoms should be prepared in a given hyperfine 
level. A microwave radiation then creates Rabi oscillations 
between this level and another hyperfine level \cite{darcy}.
These oscillations are described by the single-qubit Hamiltonian 
$\hat{H}_s$. 
Hence, the two hyperfine levels correspond to spin up and
spin down.
The kicking strength $K$ depends on the hyperfine level, 
as required in our model. 
After a sequence of kicks, the lattice is switched off and 
the momentum distribution of the cold particles can be measured.
As we will show in this paper, this can give information on 
the state (hyperfine level) of the qubit. 
We point out that a similar experiment has been proposed
\cite{zoller} and experimentally realized \cite{darcy}
with a different purpose, namely, to measure the fidelity
of quantum motion under a Hamiltonian's perturbation
(see, \textit{e.g.}, 
\cite{peres,jalabert,jacquod,benenti,prosen,zurek} 
and references therein).
Therefore, the experimental implementation
of our detector model is in principle possible, even though
the realization of the kicked rotator dynamics has been 
so far possible only for values 
$\hbar\sim 1$ or larger \cite{raizen2,amman,delande,darcy}.

\section{Dynamical decoherence}

In this section, we study the dynamical decoherence of the system, 
induced by the coupling to the detector. In this context, the detector 
can be seen as a microscopic model of a chaotic bath.  
We consider two time scales $T_1$ and $T_2$ that characterize
the relaxation of the diagonal elements of the 
system density matrix and the decay of the off-diagonal
elements, respectively. As we are interested in the measurement of 
the spin polarization along the $z$ axis, the time scales 
$T_1$ and $T_2$ will be derived in the basis of the eigenvectors
of $\hat{\sigma}_z$.
A short discussion of the time scales that characterize the
evolution of diagonal and off-diagonal elements of the
qubit's density matrix in the basis of the energy eigenstates 
of the qubit (the $\hat{\sigma}_x$ basis) will be given in 
Appendix B.
We consider the case $\delta \ll 1$, 
corresponding to a time between kicks (qubit-detector
interactions) much smaller than the period of the free-qubit
Rabi oscillations \cite{deltasmall}. 
We also derive, in the framework of the quantum operations 
formalism, a phase damping map for the system and compare the results 
of this approach with those obtained from the exact
numerical solution of the overall dynamics (system plus detector).

\begin{figure}
\centerline{\epsfxsize=8.5cm\epsffile{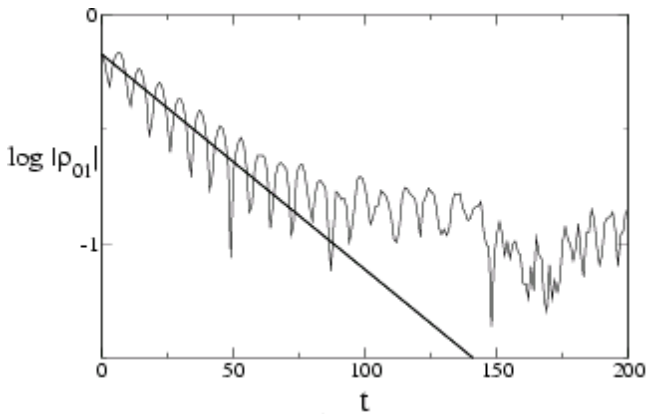}}
\caption{
Dephasing of the spin system
for $K=8$, $\epsilon=\epsilon_c/\hbar=0.3$, 
$\delta=0.2$, $\hbar=7.67\times 10^{-4}$,
and initial spin state 
$|\psi_s\rangle=(|0\rangle+2|1\rangle)/\sqrt{5}$. 
The straight line fit gives 
$|\rho_{01}|=a\exp(-\Gamma_2 t)$,
with $a=0.40$ and $\Gamma_2=2.17\times 10^{-2}$. 
Here and in the following figures the logarithms are decimal.
}
\label{fig1}       
\end{figure}

The dephasing induced by the detector is illustrated in 
Fig.~\ref{fig1}.
We can clearly see the exponential decay of the non diagonal 
term $\rho_{01}$ of the reduced density matrix 
$\rho(t)={\rm Tr}_{d}[|\Psi(t)\rangle\langle \Psi(t)|]$ 
of the two-level system (${\rm Tr}_d$ denotes the trace over
the detector). Since the overall Hilbert space is finite, 
the exponential decay is possible only up to a finite time,
after which quantum fluctuations determine the residual value
of $\rho_{01}$.
Nevertheless, the decay rate $\Gamma_2$
can be clearly extracted from a fit of the short-time 
decay 
\begin{equation}
|\rho_{01}(t)|\propto \exp(-\Gamma_2 t). 
\end{equation}
This gives us the dephasing time scale $T_2=1/\Gamma_2$.
We also note that the exponential decay is superimposed
to Rabi oscillations with period given by 
$t_R=2\pi/\omega_R=\pi/\delta$.
This is due to the fact that the Hamiltonian $H_s$ induces
free rotations of the spin around the $x$-axis with the rotation
period being $t_R$.

\begin{figure}
\centerline{\epsfxsize=8.5cm\epsffile{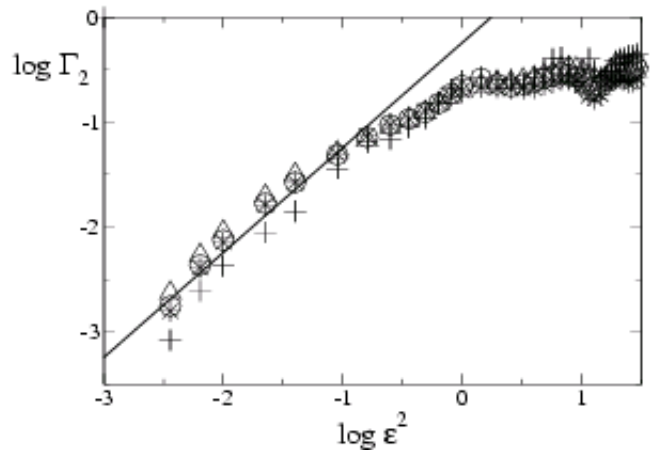}}
\caption{
Dependence of dephasing rate $\Gamma_2$ on $\epsilon^2$
for $\delta=0.1$, 
$\hbar=4.91\times 10^{-2}$ (plus), 
$\hbar=1.23\times 10^{-2}$ (triangles), 
$\hbar=3.07\times 10^{-3}$ (circles), and 
$\hbar=7.67\times 10^{-4}$ (stars).
The initial state and $K$ are the same as in Fig.~\ref{fig1}.
The straight line shows the average fit
$\Gamma_2= A \epsilon^2$, with $A=0.57$. 
}
\label{fig2}       
\end{figure}

The dependence of the decay rate $\Gamma_2$ on the coupling strength
$\epsilon=\epsilon_c/\hbar$ is given in Fig.~\ref{fig2}, for 
different values of the effective Planck constant $\hbar$.
For weak coupling $\epsilon\ll 1$, we have in average
\begin{equation}
\Gamma_2 \approx \epsilon^2/2 ,
\end{equation}
in agreement with the expectations of the Fermi golden rule.
For $\epsilon>1$, the dephasing rate saturates to an 
$\epsilon$-independent value. The discussion of the strong 
coupling regime is postponed to the end of this section. 

As we have seen in Fig.~\ref{fig1}, at long times $\rho_{01}(t)$
oscillates around a residual value $\rho_{\rm res}$.
In Fig. 3, we show the dependence of $\rho_{\rm res}$ 
on the overall size $N=N_dN_s$ of the Hilbert space,
where $N_d$ and $N_s=2$ are the dimensions of the Hilbert spaces 
for the detector and the spin, respectively.
We can see that 
\begin{equation}
\rho_{res}\propto 1/\sqrt{N}, 
\end{equation}
as expected from the following simple
statistical estimation. We assume that at long times the overall 
wave function is ergodic, that is,
\begin{equation}
|\Psi(t)\rangle=\sum_{i=0}^1\sum_{n=1}^{N_d} c_{in}(t) |i\rangle |n\rangle,
\end{equation}
where the coefficients $c_{in}(t)$ have amplitudes 
$\sim 1/\sqrt{N}$ (to assure the wave function normalization) and
random phases. Therefore, 
$\rho_{01}(t)=\sum_n c_{0n} c_{1n}^\star \sim 1/\sqrt{N}$ 
(sum of $N/2$ terms of amplitude $\sim 1/N$ with random phases).

\begin{figure}
\centerline{\epsfxsize=8.5cm\epsffile{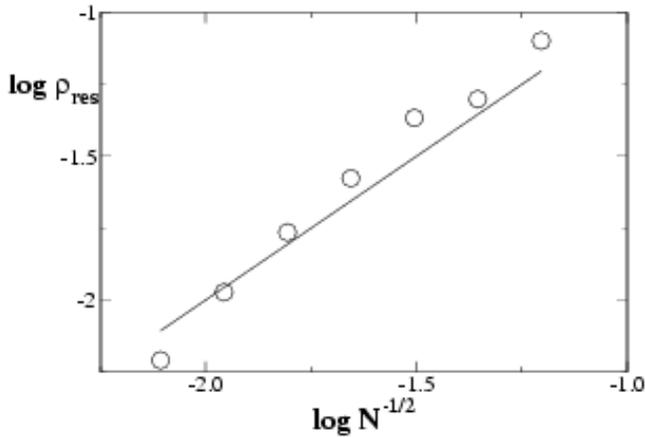}}
\caption{
Dependence of $\rho_{\rm res}$ 
(obtained after averaging $|\rho_{01}(t)|$ over the 
time interval $500\le t \le 2000$) on the size $N$
of the overall Hilbert space.
Parameter values are the same as in Fig.~\ref{fig1}.  
The straight line gives $\rho_{\rm res}= 1/\sqrt{N}$. 
}
\label{fig3}       
\end{figure}

\begin{figure}
\centerline{\epsfxsize=8.5cm\epsffile{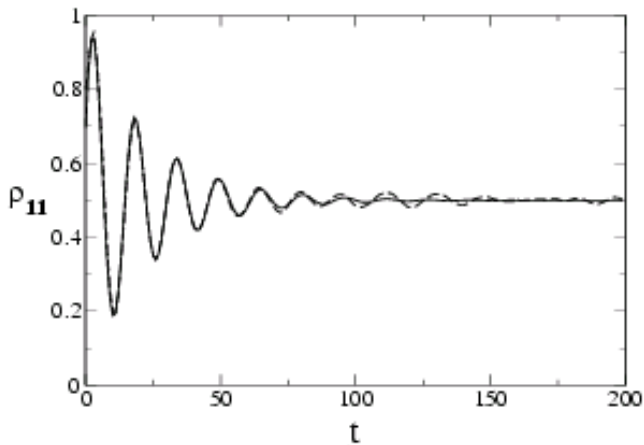}}
\caption{
Population relaxation, that is, evolution in time of 
$\rho_{11}$ (solid curve). Parameter values are as in Fig.~\ref{fig1}.
The dashed curve shows the fit $\rho_{11}=\frac{1}{2}+
a\sin(bt+\phi)\exp(-\Gamma_1 t)$,
with $a=0.5$, $b=0.404$, $\phi=0.405$ and $\Gamma_1=4.36\times 10^{-2}$. 
}
\label{fig4}       
\end{figure}

The relaxation of the diagonal component $\rho_{11}(t)$ to 
the asymptotic value $\rho_{11}=\frac{1}{2}$ is shown in Fig.~\ref{fig4}
\cite{fluctuations}. It can be seen that the population 
relaxation is exponential, with rate $\Gamma_1$. Indeed, the 
evolution in time of $\rho_{11}(t)-\frac{1}{2}$ is well fitted by the 
curve $a\sin(bt+\phi)\exp(-\Gamma_1 t)$, with 
$b=0.404\approx 2\delta= 0.4$ frequency of the free-qubit 
oscillations and $\Gamma_1=4.36\times 10^{-2}$. This fit allows 
us to extract the relaxation time scale $T_1=1/\Gamma_1=22.9$.

\begin{figure}
\vspace{0.5cm}
\centerline{\epsfxsize=8.5cm\epsffile{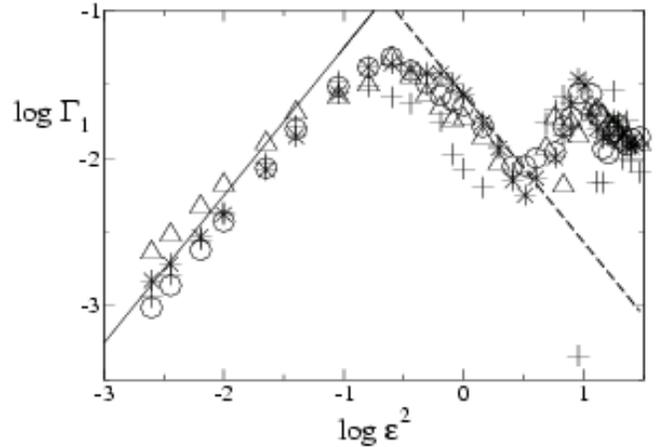}}
\caption{
Dependence of the relaxation rate $\Gamma_1$ 
on the coupling strength $\epsilon$,
at $\delta=0.1,K=8$, 
$\hbar=4.91\times 10^{-2}$ (plus), 
$\hbar=1.23\times 10^{-2}$ (triangles), 
$\hbar=3.07\times 10^{-3}$ (circles), and 
$\hbar=7.67\times 10^{-4}$ (stars).
The initial state of the qubit is 
$|\psi_s\rangle=(|0\rangle+2|1\rangle)/\sqrt{5}$. 
The straight lines show $\Gamma_1=A\epsilon^2$,
with $A=0.56$ (solid line) and $\Gamma_1=B/\epsilon^2$ 
with $B/\delta^2=2.7$ (dashed line).
}
\label{fig5}       
\end{figure}

The dependence of the relaxation rate $\Gamma_1$ on the 
coupling strength $\epsilon$ is shown in Fig.~\ref{fig5}.
At small $\epsilon$, the behavior of the relaxation rate is 
similar to that of the dephasing rate: we have 
\begin{equation}
\Gamma_1\approx \Gamma_2 \approx \epsilon^2/2.
\label{eq7}
\end{equation}
Moreover, the relaxation rate exhibits an interesting non 
monotonous behavior: it has a maximum at 
$\Gamma_1\sim \delta$ and then decays when increasing
$\epsilon$.
This phenomenon is a manifestation of the quantum Zeno
effect \cite{sudarshan,raizen3}: 
repeated measurements performed by the environment
(the detector) prevent the system from relaxing. 
In the Zeno regime, the stronger the dephasing, the slower
the relaxation. In our model, this is true in a rather broad
range of $\epsilon$, that is $\sqrt{\delta} \ll \epsilon < 1$
(see Appendix B).
In the case of Ohmic dissipation
one expects \cite{schon,schon2}
\begin{equation} 
\Gamma_1\sim \delta^2/\Gamma_2
\sim \delta^2/\epsilon^2.
\label{eq8}
\end{equation}
This is in a good agreement with our numerical data,
shown in Fig.~\ref{fig5}. In the same figure, we compare 
these data with the theoretical curve $\Gamma_1=B/\epsilon^2$,
with the fitting constant $B=2.7\times 10^{-2}$ (note that 
here $\delta^2=10^{-2}$).
For strong interactions $\epsilon>1$ the relaxation 
rate exhibits an oscillating behavior. 

It is interesting to compare the results obtained from 
our deterministic detector model with those of a quantum 
map derived for the system density matrix in the 
frame of the quantum operations formalism. 

The evolution in one period of time of the system density matrix
can most conveniently be written in the Bloch sphere representation,
in which the coordinates $x$, $y$, and $z$ are related to 
the matrix elements of ${\rho}$ as follows:
$\rho_{11}=\frac{1}{2}(1-z)$ and $\rho_{01}=\frac{1}{2}(x-iy)$.
The free evolution between two consecutive kicks is ruled by the free 
Hamiltonian $\hat{H}_{s}$, that is, 
\begin{equation}
\tilde{\rho}=e^{-i\delta\sigma_x}\rho e^{i\delta\sigma_x}.
\end{equation}
From this equation we obtain
\begin{equation}
\left\{
\begin{array}{l}
\tilde{x}=x,\\
\tilde{y}=\cos(2\delta)y-\sin(2\delta)z,\\
\tilde{z}=\sin(2\delta)y+\cos(2\delta)z,
\end{array}
\right.
\end{equation}
the coordinates $\tilde{x},\tilde{y},\tilde{z}$ corresponding
to the density matrix $\tilde{\rho}$.

We model the effect of the interaction with the detector 
as a phase kick. That is to say, we assume that in 
the interaction Hamiltonian of Eq.~(\ref{Hamiltonian}) the angle 
$\theta$ is drawn from a random uniform distribution in $[0,2\pi]$.
Therefore, the density matrix $\bar{\rho}$, obtained after 
averaging over $\theta$, is given by
\begin{equation}
\bar{\rho}=\frac{1}{2\pi}
\int_0^{2\pi} d\theta R(\theta) \tilde{\rho} R^\dagger(\theta),
\end{equation}
where
\begin{equation}
R(\theta)=\left[
\begin{array}{cc}
e^{-i\epsilon\cos\theta} & 0 \\
0 & e^{i\epsilon\cos\theta} 
\end{array}
\right].
\label{phasedamping}
\end{equation}
For $\epsilon\ll 1$, we obtain
\begin{equation}
R(\theta) \approx I - i\epsilon \cos(\theta) \sigma_z -\frac{\epsilon^2}{2}
\cos^2(\theta) I.
\end{equation}
Using this approximation, we end up with the map
\begin{equation}
\begin{array}{c}
{\displaystyle
\bar{\rho} = \tilde{\rho} 
+\epsilon^2\sigma_z\tilde{\rho} \sigma_z   
<\cos^2 \theta > - \epsilon^2 <\cos^2\theta>\tilde{\rho}} \\
{\displaystyle
= \left(1-\frac{\epsilon^2}{2}\right) \tilde{\rho} 
+\frac{\epsilon^2}{2} \sigma_z\tilde{\rho}\sigma_z},
\end{array}
\label{phasedampingmap}
\end{equation}
where in the random phase
approximation we take
$<\cos^2\theta>=\frac{1}{2\pi}\int_0^{2\pi}d\theta \cos^2\theta
=\frac{1}{2}$.
This map is known as the phase damping noise channel \cite{chuang,cpos}. 

Map (\ref{phasedampingmap}) can be written in the 
Bloch sphere coordinates as follows:
\begin{equation}
\left\{
\begin{array}{l}
\bar{x}=(1-\epsilon^2)\tilde{x}
=(1-\epsilon^2)x,\\
\bar{y}=(1-\epsilon^2)\tilde{y}
=(1-\epsilon^2)[\cos(2\delta)y-\sin(2\delta)z],\\
\bar{z}=\tilde{z}
=\sin(2\delta)y+\cos(2\delta)z,
\end{array}
\right.
\label{eq:map}
\end{equation}
where the coordinates $\bar{x},\bar{y},\bar{z}$ correspond
to $\bar{\rho}$.
An alternative derivation of map (\ref{eq:map}), based on 
the master equation approach, is provided in Appendix A.

\begin{figure}
\centerline{\epsfxsize=8.cm\epsffile{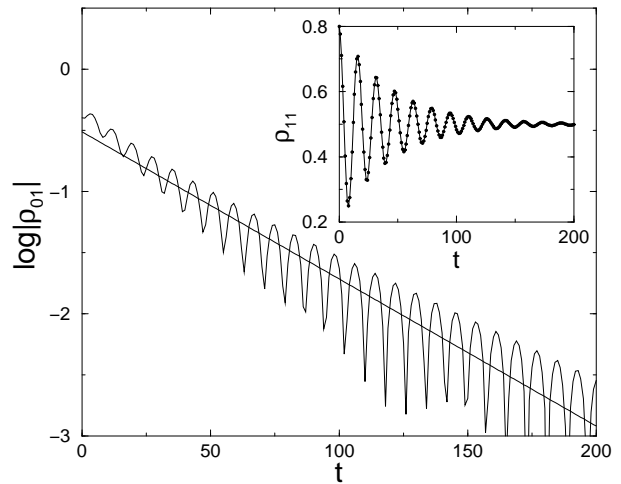}}
\caption{Dephasing and relaxation in the phase damping  
map (\ref{eq:map}). Main figure: time dependence of $|\rho_{01}|$ 
for $\epsilon=0.225$ and other parameter values as in Fig.~\ref{fig1}.
The straight line fit gives the exponential decay 
$|\rho_{01}|=a\exp(-\Gamma_2 t)$, with $a=0.31$ and 
$\Gamma_2=2.77\times 10^{-2}=0.55 \epsilon^2$.
Inset: same as in the main figure but for 
$\rho_{11}$. The behavior of $\rho_{11}$ (dots) is very well 
reproduced by the fit
$\rho_{11}=\frac{1}{2}+a\sin(bt+\phi)\exp(-\Gamma_1 t)$,
with $a=0.3$, $b=0.4$, $\phi=1.52$ and $\Gamma_1=2.30\times 10^{-2}
=0.45 \epsilon^2$ 
(solid curve).}
\label{fig6}
\end{figure}

This map gives the evolution of the Bloch sphere coordinates in 
one kick and can be iterated. From the values of the coordinates
$x$, $y$, and $z$ after $t$ map steps we can obtain 
$\rho_{01}$ and $\rho_{11}$ at time $t$. 
An example of numerical solution of map (\ref{eq:map}) is shown 
in Fig.~\ref{fig6}. The exponential decay 
of $\rho_{01}$ and the exponential relaxation of $\rho_{11}$
to the steady state value $\frac{1}{2}$ can be clearly seen. 
Similarly to Fig.~\ref{fig1} and Fig.~\ref{fig4}, we also have 
oscillations with the Rabi frequency.
We also checked that the relaxation rates $\Gamma_1$ and $\Gamma_2$
are both proportional to $\epsilon^2$. More precisely, the numerical
data for the model (\ref{eq:map}) give
$\Gamma_1\approx\Gamma_2\approx 0.53\epsilon^2$. 
These rates are in good agreement with those  from the 
numerical data obtained for the whole system (qubit plus detector)
and shown in Fig.~\ref{fig2} and Fig.~\ref{fig5}.
Therefore, it is noteworthy that our fully deterministic dynamical 
model can reproduce the main features of the single-qubit decoherence 
due to a heat bath. 
Our results are also in agreement with the analytical solution obtained 
in the continuous time limit and described in Appendix B. 

\begin{figure}
\centerline{\epsfxsize=8.5cm\epsffile{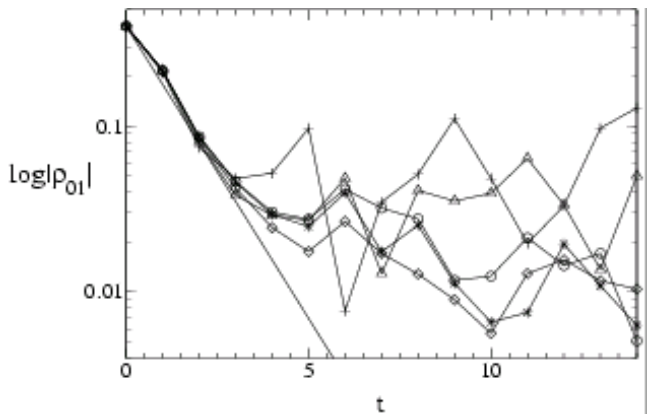}}
\caption{Time dependence of $|\rho_{01}|$ for $K=4.5$,
$\epsilon_c\equiv \epsilon 2\pi/2^{n-1}=0.8$, and $\delta=0.01$. 
The initial state of the qubit is 
$|\psi_s\rangle=(|0\rangle +2|1\rangle)/\sqrt{5}$,
the initial detector state is a Gaussian wave packet
 with area size $\hbar$
centered at $p=0$, $\theta=\pi$.
Data are shown for  
$\hbar=4.91\times 10^{-2}$ (plus), 
$\hbar=1.23\times 10^{-2}$ (triangles), 
$\hbar=3.07\times 10^{-3}$ (circles), 
$\hbar=7.67\times 10^{-4}$ (stars), and
$\hbar=1.92\times 10^{-4}$ (diamonds).
The straight line represents the exponential decay 
with rate given by the Lyapunov exponent 
$\lambda\approx \ln(K/2)=0.81$ \cite{chirikov79}. }
\label{fig7}       
\end{figure}

To close this section, we discuss the case in which the
coupling strength is classically strong, namely $\epsilon_c\sim 1$,
corresponding to $\epsilon\gg 1$ for a quasi-classical detector 
($\hbar\ll 1$). As shown in Fig.~\ref{fig7}, the dephasing rate
is given by the Lyapunov exponent $\lambda$ of the underlying classical 
chaotic dynamics of the kicked rotator. We point out that, in this
regime, the dephasing rate is independent of the interaction 
strength. This result can be understood as follows.
Assuming that the period of the free oscillations
of the system is much larger than the dephasing time, namely 
$1/\delta\gg\lambda$, we have
\begin{equation}
\rho_{01}(t)\approx \rho_{01}(0)\langle \psi_d |(\hat{U}_{d,+}^\dagger)^t
(\hat{U}_{d,-})^t | \psi_d\rangle,
\label{eq:fidelity}
\end{equation} 
where $\hat{U}_{d,+}$ and $\hat{U}_{d,-}$ are the one-kick evolution 
operators for the detector when the effective kicking strengths are
$K_{\rm eff}=K_+=K+\epsilon_c$ and 
$K_{\rm eff}=K_-=K-\epsilon_c$, respectively
\cite{rho11}.
For $K_{\rm eff}=K_-$, the initial Gaussian wave packet is 
centered at a stable fixed point and therefore, for short
times, $(\hat{U}_{d,-})^t|\psi_d\rangle\approx |\psi_d\rangle$.
On the other hand, the same fixed point is unstable 
for $K_{\rm eff}=K_+$. Therefore, in the quasi-classical 
regime and for strong enough perturbations the wave packet 
spreads along the direction of instability for the classical
motion, with rate given by the Lyapunov exponent.
We note that this phenomenon is closely related to the Lyapunov 
decay of the fidelity of quantum motion in chaotic 
systems \cite{jalabert,jacquod,benenti,prosen,zurek}. 
Indeed, Eq.~(\ref{eq:fidelity}) shows that $\rho_{01}(t)$ is 
proportional to the fidelity amplitude 
$f(t)=\langle \psi_d |(\hat{U}_{d,+}^\dagger)^t
(\hat{U}_{d,-})^t | \psi_d\rangle$. This quantity measures 
the stability of quantum motion under perturbations. More 
precisely, $f(t)$ is the overlap of two states which, 
starting from the same initial conditions, evolve 
under two slightly different Hamiltonians. 
The above discussion shows that the fidelity decay has an interesting 
interpretation in terms of dephasing of an appropriate 
two-level systems.

\section{Detector model}

In this section, we study the efficiency of the detector 
in the regimes of weak and strong system-detector coupling.

\begin{figure}
\epsffile{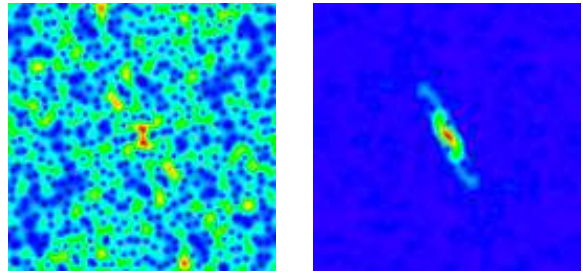}
\caption{(Color online) 
Husimi function in action-angle variables 
$(p,\theta)$ for the detector, with $-\pi\le p < \pi$ 
(vertical axis) and $0\le \theta <2 \pi$ (horizontal axis)
for the kicked rotator coupled to up spin (left) and down spin (right),
at $K=4.5$, $\epsilon_c=0.8$, $\delta=0.1$, 
$\hbar=1.23 \times 10^{-2}$, $t=20$.
The initial states of the kicked rotator and the qubit are
a Gaussian packet centered at the fixed point $p=0$, $\theta=\pi$
and $|\psi_s\rangle=(|0\rangle+|1\rangle)/\sqrt{2}$. 
Color represents the density from blue/black (minimal value) to red/gray 
(maximal value).
}
\label{fig8}       
\end{figure}

\begin{figure}
\vspace{0.5cm}
\centerline{\epsfxsize=8.5cm\epsffile{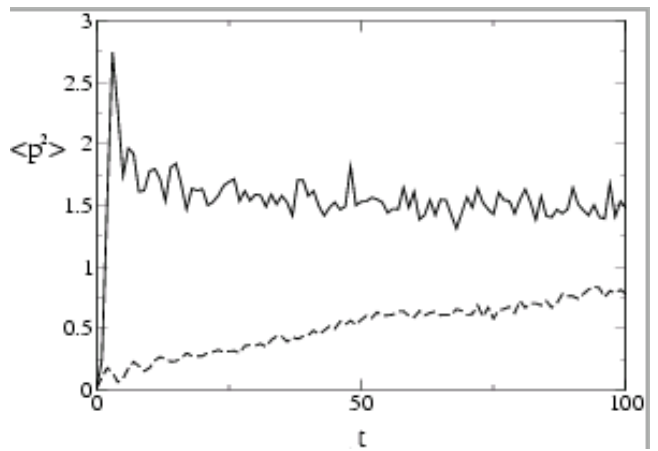}}
\caption{Time  dependence of $\langle p^2\rangle $
for up spin (curve) and down spin (dashed curve).
Parameter values are the same as in Fig.\ref{fig8}.}
\label{fig9}       
\end{figure}

Let us first discuss the strong measurement case with
$\epsilon_c$ large $\epsilon_c \sim 1$.
In this case, the effective coupling strength 
$K_{\rm eff}=K+\epsilon_c \sigma_z$ significantly depends 
on the up or down state of the spin system.
Therefore, it is easy to find regimes in which the response 
of the detector clearly depends on the state of the system. 
An example is shown in Figs.~\ref{fig8}-\ref{fig9}. In 
these figures, we have $K=4.5$, while 
$K_{\rm eff}=K-\epsilon_c=3.8$ when the rotator is coupled 
to a down spin ($\sigma_z=-1$) and  
$K_{\rm eff}=K+\epsilon_c=5.2$ when the rotator is coupled 
to an up spin ($\sigma_z=+1$). 
The initial Gaussian packet is centered at the fixed point 
$p=0$, $\theta=\pi$. A linear stability analysis of the 
classical dynamics of the kicked rotator shows that 
this fixed point is stable for $0\le K \le 4$ and unstable 
for $K>4$ and $K<0$. Therefore, in the case of coupling to a 
down spin, the fixed point is stable ($K_{\rm eff}<4$), while
the same point is unstable when the detector is coupled to an
up spin ($K_{\rm eff}>4$). 
In Fig.~\ref{fig8}, we see that, if the qubit is in its up state,
the Husimi function is spread in the phase space
after $t=20$ kicks (left plot) \cite{husimi}. 
On the contrary, if the qubit is in its down state, 
the Husimi function is localized around the stable fixed point
at $t=20$ (right plot).  
The difference between the Husimi distributions in these two cases
is evident and leads to measurable effects.
Indeed, as shown in Fig.~\ref{fig9}, the values of the second 
moment $<p^2>$ are very different at short times for up and down
spins. 

\vspace{0.5cm}

\begin{figure}[tb!]
\centerline{\epsfxsize=6.5cm\epsffile{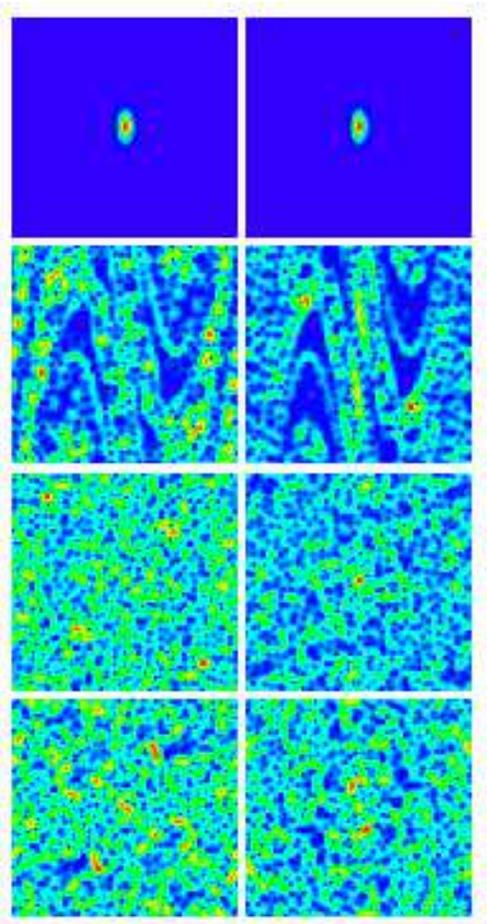}}
\caption{(Color online) Hypersensitivity of the Husimi function 
on the spin value for 
$K=8$, $\epsilon=0.4$, $\delta=0.2$, and 
$\hbar=1.23\times 10^{-2}$.
From top to bottom $t=0$, $4$, $8$, $12$.
The left plots are for up spin, the right ones for down spin.
The initial states of the kicked rotator and of the qubit are
a Gaussian packet centered at the fixed point $p=0$, $\theta=\pi$
and $|\psi_s\rangle=(|0\rangle+|1\rangle)/\sqrt{2}$. 
The color code is as in Fig.~\ref{fig8}.}
\label{fig10}       
\end{figure}

\begin{figure}[ht!]
\centerline{\epsfxsize=8.5cm\epsffile{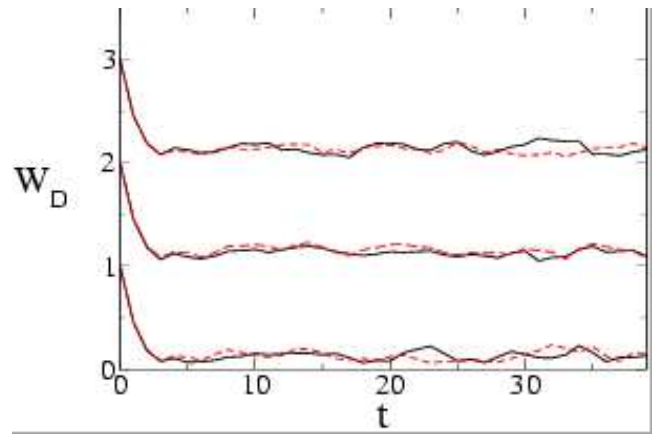}}
\caption{(Color online)
Integral $W_D$ of the Husimi distribution
over a square box of size 
$2.53\times 10^{-1}\gg \hbar=1.23\times 10^{-2}$,
centered at $p=0$, $\theta=\pi$, 
for spin up (dashed red/gray curve) and spin down
(solid black curve) states.
From top to bottom: 
($\epsilon=0.4$, $\delta=0.0$),
($\epsilon=0.2$, $\delta=0.2$), and
($\epsilon=0.4$, $\delta=0.2$).
The initial state of the detector and the other parameter
values are as in Fig.~\ref{fig10}.
The top and middle curves are shifted in the $y$-axis direction
by $+2$ and $+1$, respectively.}
\label{fig11}       
\end{figure}

It is interesting to discuss the case of weak system-detector
coupling with $\epsilon\sim 1$, 
$\epsilon_c=\epsilon\hbar\ll 1$. In the chaotic regime for the
kicked rotator the Husimi function exhibits hypersensitivity to
perturbations \cite{simone}. 
Indeed, as shown in Fig.~\ref{fig10}, we have 
markedly different Husimi functions when the detector is 
coupled to up or down spin. In the case of Fig.~\ref{fig10},
we have $\epsilon=0.4$ and $\hbar=1.23\times 10^{-2}$, 
corresponding to $\epsilon_c=4.91\times 10^{-3}\ll 1$.
However, it is unclear how to extract measurable information
from this difference. Indeed, it is reasonable to assume that 
only the coarse graining properties of the semi-classical detector 
are accessible, the size of the coarse graining being much larger
than the Planck cell.
The difficulty to distinguish the left and right Husimi plots 
of Fig.~\ref{fig10} after coarse graining is illustrated in 
Fig.~\ref{fig11}. In this latter figure we compute
the integral $W_D$ of the Husimi distribution
over a box of size $2.53\times 10^{-1}\gg \hbar=1.23\times 10^{-2}$, 
centered, as the initial wave packet, at $p=0$, $\theta=\pi$. 
The evolution in time of $W_D$ is shown for different but small
coupling strengths ($\epsilon_c\ll 1$). It is clear that the 
evolutions of $W_D(t)$ for up and down spin states can be hardly
distinguished.

\section{Conclusions}

In this paper we have proposed and studied a deterministic model 
of quasi-classical detector coupled to a single-qubit system. 
Our results show that the detector introduces dynamical decoherence 
in the system. Moreover, we have shown that an efficient measurement 
is possible in the case of strong system-detector coupling.
In the case of weak coupling, the chaotic dynamics of the detector
is still hypersensitive to the state of the qubit but it is 
unclear how to detect the up/down states in a coarse graining 
measurement.
It is an interesting question whether this conclusion remains 
valid when the unavoidable coupling of the detector with 
a dissipative environment is taken into account. 

We thank the IDRIS in Orsay and CalMiP in Toulouse for 
access to their supercomputers.
This work was supported 
in part by the 
project EDIQIP of the IST-FET program of the EC.

\appendix

\section{Master equation}

In this Appendix, we provide an alternative derivation of
the phase damping map (\ref{eq:map}), based on 
the master equation approach. For this purpose, we start from 
the overall Hamiltonian of Eq.~(\ref{Hamiltonian})
and make the usual Born and Markov approximations. 
That is to say, we assume that the detector's dynamics is 
practically unaffected by the interaction with the qubit 
and that any effect  the qubit has on the detector 
is limited to a time scale much shorter than the time 
scales of interest for the dynamics of the qubit.
The first condition is fulfilled in the case of weak 
coupling, $\epsilon\ll 1$, the second is satisfied 
due to the fast decay of correlations for the detector 
in the chaotic regime (here we assume that $\delta\ll 1$ and
that correlations decay in a few kicks).
Under these assumptions, we can derive \cite{master} the following
master equation:
\begin{equation}
\begin{array}{c}
\displaystyle{
\frac{d \hat{\rho}}{dt}=-i[\hat{H}_s,\hat{\rho}] +
\frac{\gamma}{2} \left (
[\hat{\sigma}_z, \hat{\rho} \hat{\sigma}_z]+
[\hat{\sigma}_z \hat{\rho}, \hat{\sigma}_z]
\right)}\\
\displaystyle{
=-i[\hat{H}_s,\hat{\rho}] + \frac{\epsilon_c^2}{2} 
(\hat{\sigma}_z\hat{\rho}\hat{\sigma}_z-\hat{\rho})
\sum_m\delta(t-m),
}
\end{array}
\label{eq:lindblad}
\end{equation}
where we have used  
\begin{equation}
\begin{array}{c}
{\displaystyle \gamma = \epsilon_c^2 
{\rm Tr}_d[\cos^2 (\tilde{\theta}) \tilde{\rho}_d]
\sum_m \delta (t-m)
}\\
{\displaystyle =\frac{\epsilon_ c^2}{2} 
\sum_m \delta (t-m).}
\end{array}
\label{eq10}
\end{equation}
Here $\tilde{\rho}_d$ is the detector's density matrix and
the tilde denotes the fact that we are using the interaction picture
(that is, the time evolution of the operators with a tilde 
is ruled by the detector's Hamiltonian $\hat{H}_d$).
Assuming that the density matrix $\tilde{\rho}_d$ corresponds 
to a uniform distribution in the $\theta$ variable, we obtain 
${\rm Tr}_d[\cos^2 (\tilde{\theta}) \tilde{\rho}_d]
\approx\frac{1}{2\pi}\int_0^{2\pi} d\theta\cos^2\theta=\frac{1}{2}$.
The integration of the master equation (\ref{eq:lindblad}) in one time
step leads to the phase damping map (\ref{eq:map}).

\section{Continuous model}

The continuous version of the phase damping map (\ref{eq:map}) is
\begin{equation}
\left\{ \begin{array}{l}
\dot{x}=-\epsilon^2 x, \\
\dot{y}=-\epsilon^2 y - 2\delta z,\\
\dot{z}= 2\delta y . \end{array} \right.
\label{cont} \end{equation}

The solution for $x(t)$ is a simple decay with the rate 
$\Gamma=\epsilon^2$:
\begin{equation}
x(t) = x(0) e^{-\Gamma t} .
\end{equation}
The solution for $y(t)$ and $z(t)$ reads as follows:
\begin{equation}
\left\{
\begin{array}{l}
y(t) = [a \sin (\omega t) +b \cos (\omega t)]e^{-\Gamma t/2},\\
z(t) = [c \sin (\omega t) +d \cos (\omega t)]e^{-\Gamma t/2},
\end{array}
\right.
\end{equation}
where
\begin{equation}
\omega = \sqrt{4\delta^2 - \frac{\Gamma^2}{4}}
\end{equation} 
and the coefficients $a,b,c,d$ can be expressed as a function 
of the initial conditions $y(0),z(0)$.
Therefore, for $\epsilon\ll 2\sqrt{\delta}$, in the 
$\hat{\sigma}_z$-basis both the diagonal and the off-diagonal elements 
of the density matrix decay as $\frac{\Gamma}{2}$, in agreement 
with our numerical data shown in Sec.~III. Indeed, we have 
at large times $|\rho_{11}-\frac{1}{2}|=|z|\sim \exp(-\Gamma t/2)$ and  
$|\rho_{01}|=\frac{1}{2}(x^2+y^2)^{1/2}\sim |y|\sim \exp(-\Gamma t/2)$.

When the coupling to the detector becomes strong, so that
$\Gamma>4\delta$, then the oscillations in $y(t)$ and $z(t)$ turn into
decay with two characteristic rates:
\begin{equation}
\Gamma_{\pm} = \frac{\Gamma}{2} \pm 
\sqrt{\frac{\Gamma^2}{4} -4\delta^2} \; .
\end{equation}
The smallest rate  ($\Gamma_-$) describes the quantum Zeno effect. 
For $\Gamma\gg 4\delta$ we obtain 
$\Gamma_- \simeq 4\delta^2/\Gamma$, in agreement with 
our numerical results shown in Fig.~\ref{fig5}.

It is important to remark that in the basis of the eigenstates
of the Hamiltonian $\hat{H}_s$ one can recover the relaxation 
and decoherence time scales usually discussed in the literature.
In this basis $x$ describes the deviation of the diagonal 
elements of the density matrix from $1/2$, while $y$ and $z$
give the decay of the real and imaginary part of the off-diagonal
matrix elements. Therefore, for $\epsilon< 2\sqrt{\delta}$ 
the diagonal elements decay with rate $\Gamma$ and the 
off-diagonal elements with rate $\Gamma/2$.


\begin{thebibliography}{99}
\bibitem{deltasmall}
We have checked that the times $T_1$ and
$T_2$ (measured in number of kicks) do not significantly depend on 
$\delta$, provided that $\delta<0.5$.
\bibitem{cpos}
Map (\ref{phasedampingmap}) is completely positive and can be  
written in the Kraus form as follows: 
$\bar{\rho}=M_0\tilde{\rho} M_0^\dagger+M_1\tilde{\rho}M_1^\dagger$,
with the Kraus operators $M_0=\sqrt{1-\frac{\epsilon^2}{2}}I$ and
$M_1=\frac{\epsilon}{\sqrt{2}}\sigma_z$ satisfying the 
condition $M_0M_0^\dagger+M_1M_1^\dagger=I$.
\bibitem{rho11}
The same argument tells us that, at short times, populations
essentially do not decay, that is, $\rho_{11}(t)\approx \rho_{11}(0)$
for $t\ll 1/\delta$.
\end{thebibliography}
\end{document}